\documentclass[12pt, preprint]{aastex}
\usepackage{natbib}
\usepackage{amsmath}
\usepackage{graphicx}
\usepackage{graphicx,epsf}
\usepackage{booktabs}
\usepackage{wasysym}
\usepackage{float}
\usepackage{rotating}
\usepackage{color}

\shorttitle{Gravitational Waves from Gamma-Ray Pulsar Glitches}

\shortauthors{Stopnitzky \& Profumo}

\newcommand{\figref}[1]{Fig.~\ref{fig:#1}}

\begin{document}

\title{Gravitational Waves from Gamma-Ray Pulsar Glitches}

\author{Elan Stopnitzky\altaffilmark{1,2} and Stefano Profumo\altaffilmark{2,3}}

\altaffiltext{1}{Department of Physics, University of Hawaii at Manoa, 2505 Correa Rd., Honolulu, HI 96822, USA}
\altaffiltext{2}{Department of Physics,  University of California, 1156 High St., Santa Cruz, CA 95064, USA}
\altaffiltext{3}{Santa Cruz Institute for Particle Physics,  University of California, 1156 High St., Santa Cruz, CA 95064, USA}

\begin{abstract}
We use data from pulsar gamma-ray glitches recorded by the Fermi Large Area Telescope as input to theoretical models of gravitational wave signals the glitches might generate. We find that the typical peak amplitude of the gravity wave signal from gamma-ray pulsar glitches lies between $10^{-23}$ and $10^{-35}$ in dimensionless units, with peak frequencies in the range of 1 to 1000 Hz, depending on the model. We estimate the signal-to-noise for all gamma-ray glitches, and discuss detectability with current gravity wave detectors. Our results indicate that the strongest predicted signals are potentially within reach of current detectors, and that pulsar gamma-ray glitches are promising targets for gravity wave searches by current and next-generation detectors.
\end{abstract}

\section{Introduction}
Direct observation of gravitational waves has so far proved to be an elusive task, albeit indirect evidence has been inferred via the orbital energy loss of binary neutron stars \citep{taylor1982new}. Several mechanisms have been identified that could make rotating neutron stars (also known as pulsars) promising candidate gravitational waves sources \citep{PhysRevD.20.351, 1996A&A...312..675B, PhysRevLett.76.352}. As a result,  pulsars, and more specifically radio-emitting pulsars, have been systematically targeted for more than a decade by gravitational wave detectors such as LIGO in the search for a signal \citep{PhysRevLett.94.181103, 2010ApJ...713..671A}.

One potential source of gravitational radiation from pulsars is a timing irregularity known as a ``glitch''. Pulsar glitches are characterized by sudden changes in rotational frequency, or ``spin'' \citep{Espinoza:2011pq} over a baseline pre-glitch frequency.  Pulsar glitches are thought to be due to instabilities that may source gravitational waves with amplitudes spanning a wide range, in dimensionless units from $10^{-23}$ to $10^{-35}$, with frequencies on the order of 1 to 1000 Hz \citep{2001PhRvL..87x1101A}. Frequencies in the kilohertz range happen to be those LIGO is most sensitive to \citep[see e.g. fig 4. of][]{Smith:2009bx}.

The detailed physical mechanism responsible for pulsar glitches is not known, and different mechanisms could be responsible for glitches observed from different pulsars \citep[see e.g.][]{Abadie:2010sf}. In younger pulsars, the mechanism thought to drive the glitch is a so-called starquake \citep{Pizzochero:2011dd}. In this scenario, as the pulsar steadily spins down, a strain is induced in the solid crust as the object tries to settle into a more spherical shape, eventually leading to a fracture in the crust and to energy release \citep{Abadie:2010sf}. 

However, the observed frequency and magnitude of glitches indicate that this mechanism cannot be responsible for glitches observed in all pulsars \citep{Abadie:2010sf}. Rather, for older and less seismically active pulsars, glitches are thought to be driven by interactions between the crust and superfluid interiors of the neutron star  \citep{Abadie:2010sf}. The magnetic field that sources the observed radio pulses is coupled to the crust and to the normal fluid interior, but the superfluid core of the pulsar may rotate with a different, unobservable velocity. Moreover, the rotation of the crust and normal fluid components slows down because of the emission of electromagnetic radiation, while the superfluid core maintains its angular velocity. The superfluid is weakly coupled to the normal components of the star, and ultimately a maximal difference in angular velocity is reached, after which the core suddenly transfers some of its reserve angular momentum to the crust, producing the glitch \citep{0264-9381-25-22-225020}. 

{Among glitches observed in radio, the fractional change in spin frequency spans seven orders of magnitude in the total population, and as many as four orders of magnitude in a single object \citep{2008ApJ...672.1103M}. Moreover, although most active pulsars have only been observed to glitch once, the majority of those that have glitched repeatedly do so at unpredictable intervals. These features suggest an avalanche mechanism responsible for glitches, in which the seismic activity or vortex unpinning may be a local or global event. In this case, the sizes and waiting times for glitches would follow power law distributions whose exponents are related. This model was explored in detail by \citet{2008ApJ...672.1103M}. In the event that glitches are indeed described as an avalanche process, the size distribution should have probability density function \citep{2008ApJ...672.1103M}:}

\begin{equation}
p \left( \frac{\Delta\nu}{\nu} \right) \propto \left( \frac{\Delta\nu}{\nu} \right)^{-a}.
\end{equation}

{This model is consistent with the data, provided that the exponent $a$ differs from pulsar to pulsar. It is observed that large glitches do not appear preferentially in any one class of object, but may occur more or less frequently in an individual pulsar \citep{2008ApJ...672.1103M}. According to the model, the waiting times themselves also follow a power law distribution, with}

\begin{equation}
p \left( \lambda , \Delta t \right)=\lambda e^{\left( -\lambda\Delta t \right)},
\end{equation}

{ where $\lambda$ is the mean glitching rate and once again varies between pulsars. Interestingly, attempts to find a correlation between the mean glitch rate and the size of the next glitch have proven unsuccessful; this fact is also consistent with the avalanche model \citep{2008ApJ...672.1103M}. The authors conclude that it is likely that all pulsars are capable of glitching, but that $\lambda$ in too small in most cases to lead to a detection over the 40-year period that they have been monitored.}

In any event, the glitch may excite oscillations or flows in the pulsar that, in turn, could source gravitational radiation  \citep{Abadie:2010sf}. The fact that all pulsars appear capable of large glitches is thus especially promising to gravitational wave detection. Depending on the nature of the source, gravitational wave signals are predicted to last anywhere from milliseconds to months, with a wide range of predicted amplitudes \citep{Prix:2011qv}.

Thus far, gravity wave searches from pulsar glitches have been triggered by radio observations only \citep{Abadie:2010sf}. Recently, however, the Fermi gamma-ray Large-Area Telescope (LAT) \citep{Atwood:2009ez} has revolutionized our understanding of the high-energy sky and of many galactic and extragalactic sources, including pulsars. Blind searches for pulsation in the gamma-ray sky, for example, have led to the discovery of many previously unknown, radio-quiet pulsars \citep[see e.g.][]{Collaboration:2010en, Pletsch:2011kp, 2010ApJ...725..571S, 2011ApJS..194...17R, 2009ApJ...706.1331A, 2041-8205-755-1-L20}. The sheer existence of several instances of ``gamma-ray only'' pulsars is  evidence that the gamma-ray beam is much wider than (or mis-aligned with) the radio beam. Importantly, a majority of the pulsars discovered by LAT are radio-quiet or extremely faint, and it is now known that these radio-quiet pulsars constitute a considerable fraction of the pulsars in the galaxy \citep{2013IAUS..291...87G}. 

Glitches from gamma-ray pulsars have also been observed and reported \citep[see e.g.][]{Ray:2010ws, 2011AAS...21743405D, fermisymp, Ray:2012ue, 2041-8205-755-1-L20}. Given the abundance of radio-quiet pulsars, it is clear that a comprehensive understanding of glitches and the gravitational waves they may produce would be incomplete without the inclusion of these sources. Furthermore, gamma-ray triggered glitches are attractive candidates because gamma-ray pulsars must typically be within a few kpc in order to be detected \citep{2013IAUS..291...81S}. LAT studies indicate that, while all glitching gamma-ray pulsars are highly energetic ($\dot E>5\times10^{35} {\rm erg\ s}^{-1}$), the size of the glitches detected in gamma-rays is comparable to that detected in radio pulsars. The proximity to the glitching neutron star imposed by requiring the event to be observable in gamma rays, and the fact that many gamma-ray pulsars are radio quiet, might indicate that some of the most promising events from the standpoint of gravity wave signals could correspond to gamma-ray glitches.

For all of these reasons, we believe it is a timely task to assess the gravity wave signal expected from gamma-ray pulsar glitches and, potentially, to follow up with dedicated searches in LIGO archival data, if feasible given the glitches' locations and times. 

A central issue in this task is the fact that gamma-ray pulsars often have poor distance determinations, especially if radio follow-ups  or follow-ups at other frequencies have not been useful to provide additional information to the gamma-ray only detection \citep{Ray:2012ue}. Often, the frequency and frequency derivative determinations, especially around a glitch, are highly uncertain. Finally, there exist significant theoretical uncertainty in the gravity wave yield that a pulsar glitch would produce, as we detail in the following sections.

In the present study, we employ a publicly available catalogue of gamma-ray pulsar glitches \citep[from][]{2011AAS...21743405D, fermisymp, 2041-8205-755-1-L20}) obtained with data from the Fermi LAT, and we make estimates for the gravity wave amplitude and peak frequency, as well as for the signal-to-noise, in various (namely 5) theoretical models for gravity wave production from neutron star glitches. As a result, we provide a list of target locations in time and in the sky where a signal for gravity waves might be detected, and we provide proof of principle that gamma-ray observations of pulsars could provide important information for gravity wave astronomy, especially for future detectors such as Advanced LIGO. We note that we show here how gamma-ray glitches are potentially promising candidate events for gravity wave searches, in a similar way that radio glitches are. We are thus not claiming that one class is more or less promising than the other but, rather, that both deserve attention in the arena of gravity wave searches.

It is important to note that besides the detected gamma-ray glitches, a substantial collective gravitational-wave emission might stem from smaller glitches as well. For example, as mentioned above \citet{2008ApJ...672.1103M} showed that the size distribution for radio pulsar glitches is consistent with being associated to a scale-invariant ``avalanche'' process. A study similar to \citet{2008ApJ...672.1103M} is however currently not possible with available gamma-ray glitches data, where only two pulsars have more than one detected glitch. We leave it to future work to address the question of the cumulative gravitational wave signal potentially associated with smaller glitches, both at radio and gamma-ray frequencies.

The remainder of this paper is organized as follows: in the following section \ref{sec:models} we outline the five models for gravity wave production from neutron star glitches; section \ref{sec:methods} presents in detail our methods and results; the final section \ref{sec:discussion} contains a discussion of our results and presents our conclusions.

\section{Glitch Models}\label{sec:models}

In this study, we consider five different models for gravitational wave production from pulsar glitches. Two of the models  analytically predict the gravitational wave signals generated by the response of the fluid interior of the pulsar to the glitch, with one model, from \citet{0264-9381-25-22-225020}, focusing on the mass quadrupole contribution to the signal (we indicate this model with the acronym ``MQ''), while the other on the current quadrupole contribution \citep{MNR:MNR17416} (we indicate this model with the acronym ``CQ''). Two additional models consider the maximal possible gravitational wave amplitude sourced by $f$-mode oscillations on the simple basis of energy conservation, with the $f$-mode oscillations driven by a starquake in one case (``FQ'' model) and by superfluid interactions in the other \citep{Abadie:2010sf} (``FS'' model). In addition to these $f$-mode oscillations, we investigate the possibility of detecting gravitational wave emission from quasi-radial oscillations excited during a glitch (``QO''), as explored in \citet{springerlink:10.1023/B:ASYS.0000003260.80468.69}. We give here a brief overview of the five models we consider in what follows.

\citet{0264-9381-25-22-225020} calculated analytically the gravitational wave signal generated in a toy-model of a pulsar glitch. They idealize the pulsar as a fluid-filled cylinder that experiences a step increase in its angular velocity, and solve for the dynamical response of the fluid to this change. For tractability, they approximate the fluid within the cylinder as having a uniform density, viscosity, and gravitational acceleration. In addition to these simplifications, a real pulsar would not exhibit such a sharp transition between fluid and crust, as the properties of the material within the pulsar would vary in a continuous fashion with depth. Their treatment considers only the mass quadrupole contribution to the gravitational wave signal, and the analysis is conducted for observers oriented alternately along the poles and along the equator of the pulsar. In spite of these simplifications, their model demonstrates how the physics of the pulsar interior may be elucidated by the features of a gravitational wave signal, and provides a rough estimate for the kind of signal that might be created. In particular, they find that the width and amplitude of the spectral peak contains information about the compressibility and stratification length-scale of the fluid. They estimate the gravitational wave characteristic amplitude, hereafter referred to by the acronym MQ for this model, with
\begin{equation}
h_{0}=\left( \frac{\delta\Omega}{\Omega} \right) \left( \frac{\pi\rho_{0}\Omega^4GL^6}{c^4gd} \right),
\end{equation}
where $\Omega$ is the angular velocity of the cylinder, $\delta\Omega$ is the difference in the angular velocity brought on by the glitch, $\rho_{0}$ is the density, the cylinder has radius $L$ and height $2L$, $c$ is the speed of light, $g$ is the gravitational acceleration, $G$ is the gravitational constant, and $d$ is the pulsar distance. In our analysis, $g$ is estimated to be $\frac{GM}{R^{2}}$, as in \citet{0264-9381-25-22-225020}. They find that for a polar observer, the ``plus'' polarization mode signal is twice as strong as that for an equatorial observer. For the cross polarization, the situation is exactly reversed. Each of the signals has angular frequency $2\Omega$, where $\Omega$ is the angular frequency of the pulsar, except for the cross polarization observed at the equator, which has angular frequency $\Omega$. These considerations are important because the sensitivity of gravitational wave detectors is frequency-dependent. For the gamma-ray pulsars we are considering here, the relevant frequency for gravitational wave signals ranges between 3 and 40 Hz. 

A follow-up study by \citet{MNR:MNR17416} examined the current quadrupole contribution to the gravitational wave signal, and generalized the analysis to include observers at arbitrary inclinations with respect to the pulsar. This model will be referred to as CQ. Their calculations indicate that, in addition to compressibility and stratification, the viscosity and inclination of the pulsar may also be inferred from the gravitational wave signal. They find that the characteristic amplitude due to the current quadrupole contribution is
\begin{equation}
h_{0}=\frac{4\pi G\rho_{0}L^6(\delta\Omega)\Omega^2}{3c^5d}.
\end{equation}
This turns out to be a stronger signal than that due to the time-varying mass quadrupole. Once again, there exist signals at both $\Omega$ and $2\Omega$, with the strength of each signal de\-pen\-ding on the inclination of the pulsar. These signals may persist for days or weeks following a glitch. 

In addition to the wave-producing flows described previously, it has been proposed that the glitch of a pulsar may excite various oscillations that are damped by gravitational wave emission \citep{MNR:MNR16497}. These oscillations would be associated with the emission of a gravitational wave signal in the form of a decaying sinusoid \citep{Abadie:2010sf}. $f$-mode oscillations, in particular, are thought to be the primary emitters of gravitational waves \citep{Abadie:2010sf}. For gravitational wave signals of this type, the LIGO  collaboration has calculated  the peak characteristic amplitude on the simple basis of energy conservation \citep{Abadie:2010sf}. We reproduce their arguments and estimates below.

For a star-quake driven glitch (FQ), the maximum possible gravitational wave amplitude may be derived by assuming that the change in energy associated with a glitch is fully radiated by gravitational waves of the form described above. When such a glitch occurs, the moment of inertia of the star changes as it settles into a more spherical shape, and its angular velocity is increased by an amount $\Delta\Omega$. Then with conservation of angular momentum, the change in rotational kinetic energy of a star with moment of intertia $I_{\ast}$ is

\begin{equation}
\Delta E_{quake}=\frac{1}{2}I_{\ast} (\Omega)(\Delta\Omega).
\end{equation}
Alternately, the glitch may be precipitated by crust/superfluid interactions (FS). In this case, the energy associated with the glitch is given by \citep{Abadie:2010sf}

\begin{equation}
\Delta E_{fluid} \approx I_{c}(\Delta\Omega)(\Delta\Omega_{lag}),
\end{equation}
where $\Delta\Omega_{lag}$ is the critical difference in rotation between crust and superfluid needed to drive a glitch ($\frac{\Delta\Omega_{lag}}{\Omega} \approx 5\times10^{-4}$), $I_{c}$ is the moment of inertia of the crust ($I_{c} \approx \frac{I_{\ast}}{10}$), and it is assumed that corotation between crust and superfluid is restored following the glitch. { It is worth cautioning, however, that in light of the avalanche model, this condition will only be met by those pulsars that glitch quasi-periodically; otherwise, the necessity of such a global event would correspond solely with the largest possible glitches \citep{2008ApJ...672.1103M}}. If all of the available energy is absorbed into the excitation of the $l=2$ spherical harmonic index of the oscillation, these energies correspond to a gravitational wave signal with peak characteristic amplitude \citep{Abadie:2010sf}

\begin{equation}
h_{0} \approx \left( \frac{2.8\times10^{-20}\ {\rm (s/kg)}^{1/2}}{\nu_{0} d} \right) \left( \frac{\Delta E}{\tau_{0}} \right)^{\frac{1}{2}},
\end{equation}
where $\nu_{0}$ is the frequency of the signal, $\tau_{0}$ is the damping time, and $d$ is the distance measured in meters. Such gravitational waves are expected to have frequencies in the range of 1-3 kHz with damping times between 0.05 and .5 seconds \citep{Abadie:2010sf}. The true amplitudes observed at Earth will in general differ from $h_{0}$, as they are modulated by a factor that depends on the polarization of the signal, inclination of the pulsar, and $m$-index of the oscillation. It is possible in principle, however, that the observed amplitude may be up to twice that of $h_{0}$. For the dependence of the amplitude on these variables, see Table 2 of \citet{Abadie:2010sf}.

We also note that burst emission from vortex avalanches following a glitch were studied in \citet{2012MNRAS.423.2058W}. In particular, \citet{2012MNRAS.423.2058W} calculate from first principles the current-quadrupole gravitational-wave signal. The signal predicted scales with the square root of the glitch size for homogeneous glitches and is proportional to the glitch size for inhomogeneous glitches. This model is closely related to the FQ and FS models we consider here, and we thus do not include it explicitly in our list of predictions.

Finally, another potential source of gravitational radiation is a quasi-radial oscillation (QO), as explored by \citet{springerlink:10.1023/B:ASYS.0000003260.80468.69}. If it is assumed that the energy from the glitch is fully converted into exciting these oscillations, the amplitude is given by 

\begin{equation}
h_{0}=\frac{4\times10^{-18}}{\omega d}(I\Omega\Delta\dot{\Omega})^{\frac{1}{2}}.
\end{equation}
Based on the equation of state assumed in this paper, such oscillations would have angular frequencies $\omega\approx 5$ kHz and would persist continuously between glitches. 

We note that the various models we consider here feature drastically different signal durations. For example, the MQ and CQ models predict signals that last from days to weeks, while the FS and FQ models have damping times between 0.05 and 0.5 seconds (thus effectively producing a burst-like signal). The QO model predicts a persistent signal between glitches. For detectors like LIGO, the effective amplitude to which the instrument is sensitive to approximately scales with the square-root of the number of wave cycles, and thus longer signals are easier to detect in general than burst-like signals. In our estimates of the signal-to-noise presented below, we consider signal durations of 0.05 to 1 s for the FS and FQ models, and of 1 to 20 days for the CQ, MQ and QO models. We also assume, below, that the observation time is always longer than the glitch duration for all models.

Searches for gravity waves from pulsars have been carried out thus far both in continuous wave mode \citep[see especially][]{2010ApJ...713..671A} and in burst-like mode \citep[see e.g.][]{Abadie:2010sf}. Here, especially for models FQ and FS, but also for the MQ and CQ models, an ``intermediate'' strategy, such as that envisioned and described in Prix et al 2011, would be highly desirable and effective at transient signals that would extend the continuous-wave signal strategy search by a finite start-time and duration.

\section{Methods and Results}\label{sec:methods}

In this section, we start with estimates of the gravitational wave amplitudes for the 5 models under consideration; we then provide a calculation of the signal-to-noise associated with all glitches and for all models, and discuss the ensuing potential for detectability; finally, we discuss in detail the impact of distance uncertainties on our conclusions.

To obtain estimates for the gravitational wave amplitudes associated with pulsar glitches detected in LAT gamma-ray data, we  assume the following fiducial pulsar parameters: $I_{\ast}=10^{38}$ kg\textperiodcentered m$^{2}$, $\rho_{0}=10^{18}$ kg/${\rm m}^{3}$, and $r=10$ km \citep{Abadie:2010sf}. For the $f$-mode oscillations, we utilize a frequency $\nu_{0}$ of $2$ kHz and a damping time $\tau_{0}$ of $0.2$ s. We note that these reference values are the same as those used in \citet{Abadie:2010sf}. The QO model requires an estimate of the fractional change in the frequency derivative during the glitch, which we have estimated to be about 0.005, based on the average of those measured for detected radio glitches in \citet{YuFebruary}. The remaining quantities necessary to produce the estimates are the pulsar spin angular frequency $\Omega$, the steady state loss in rotation $\dot{\Omega}$, the fractional change in spin frequency during a glitch $(\frac{\Delta\Omega}{\Omega})$, and the distance to the pulsar $d$. Except for the distances, we utilize data presented in \citet{fermisymp}, \citet{2041-8205-755-1-L20}, and \citet{2011AAS...21743405D} for 16 recorded glitches in 14 different gamma-ray pulsars. The distances are consolidated from \citet{1475-7516-2010-02-016}, \citet{0067-0049-187-2-460}, \citet{0004-637X-725-1-571}, \citet{2013IAUS..291..546W}, and \citet{1674-4527-11-7-007}, and are listed in Table 1 together with the pulsars' frequency, spin-down rate, fractional change in frequency during the glitch and glitch time. The last column also indicates whether or not the glitch is detected at radio frequencies. Our predictions for the gravity wave signals for all five theoretical models, and the corresponding peak frequencies, are collected in Table 2. Accurate predictions of gravitational wave amplitudes are of course systematically complicated by the often substantial uncertainty in pulsar distance.  

The significant spread in predicted values among the first four models can be explained by the very different mechanisms for gravitational wave production that each of them considers, as well as the differing assumptions they incorporate. In particular; FQ, FS, and QO assume that the energy supplied by the glitch is fully translated into exciting the various oscillations. Moreover, the signals generated in CQ, MQ, and QO are of a much longer duration than those of FQ and FS, which means they must be correspondingly weaker in amplitude, due to energy conservation.

Bearing these differences in mind, estimating the uncertainties in our predictions is a highly non-trivial task. For example, for the CQ and MQ models one would have to assess how the various assumptions employed in the toy model, i.e. cylindrical geometry and sharp boundary layer, affect the outcome expected from the actual configuration. For the remaining models, it is unclear how realistic the assumption that the energy supplied by the glitch is fully radiated in the form of gravitational waves is, as the internal physics of pulsars is not understood well enough to determine what fraction of the energy would be dissipated via different mechanisms. Nevertheless, we shall attempt below to at least estimate ``non-systematic'' uncertainties arising from possible deviations in the assumed pulsar parameters.

The strength of the MQ signal depends on the radius, density, and gravitational acceleration of the pulsar. The latter two quantities are assumed to be constant. Under the assumption that the cylindrical model employed serves as an approximation to what is in fact a spherical star, we can dispose of the dependence on the density and gravitational acceleration by writing $\rho=\frac{3M}{4\pi r^{3}}$ and $g=\frac{GM}{r^{2}}$. These assumptions ought to be more or less reasonable, since even rapidly rotating pulsars are still well approximated as spheres \citep{2007ASSL..326.....H}, and the density of a neutron star is roughly constant up to the location of the crust \citep{2007ASSL..326.....H}. In this case we are left with an equation whose dependence on the structure of the pulsar goes as $r^{5}$. Since neutron star radii are though to vary from 9 to 14 km \citep{2007ASSL..326.....H}, this introduces a fractional uncertainty in the GW amplitude of $1.1$. We note that some of the relevant parameters entering this fractional uncertainty evaluation can be inferred directly from observations of glitch recovery at radio frequencies, as described in \citet{2010MNRAS.409.1253V}.

The CQ model depends both on the density (again assumed to be uniform) and on the radius of the pulsar. Central densities may vary from those of heavy nuclei at $\rho_{0}=2.8\times10^{14}$ g\textperiodcentered cm$^{-3}$ up to possibly $15\rho_{0}$ \citep{2007ASSL..326.....H}. Central density and radius are correlated, and will be related to one another through an equation of state. However, we can bound the uncertainty by treating the two quantities separately. For this model, $h_{0}$ goes as $r^{6}$. These factors then make the amplitude for this signal uncertain up to a factor of about 1.7.

In the case of the QO model, uncertainties are due to the moment of inertia $I$, frequency of oscillations $\omega$, and fractional change in frequency derivative $\frac{\Delta \dot{\Omega}}{\dot{\Omega}}$. For stellar masses lying in the range of $1-2$ solar masses, the moment of inertia may range from about $5\times10^{37}$ to $2.5\times10^{38}$ kg\textperiodcentered m$^{2}$, depending on the equation of state \citep{2007ASSL..326.....H}. The frequency of these oscillations has little dependence on the stellar mass (and therefore on the moment of inertia) and ranges from $2-10$ kHz \citep{springerlink:10.1023/B:ASYS.0000003260.80468.69}. The fractional change in frequency derivative is estimated from the average of those measured in radio glitches in \citet{YuFebruary}. This set has an average of $4.84\times10^{-3}$ and an SDM of $0.696\times10^{-3}$. Combining these factors, we find that this estimate is uncertain up to a factor of roughly 2. 

Uncertainties in the FS and FQ models are due to the moment of inertia, the fraction of moment of inertia contained in the crust, frequency of the signal, and characteristic damping time. The crustal moment of inertia is bounded from below at about $1.5\%$ of the total \citep{2007ASSL..326.....H} and may range up to $10\%$ \citep{Abadie:2010sf}. The characteristic time-scale is related, in general, to the mean density of the mass involved. Numerical results \citep{1998MNRAS.299.1059A} indicate that the relation between f-mode frequencies and the mean density is close to linear, and that it scales with the mean density as $\omega_f\sim\bar \rho^{1/2}$. This scaling allows for estimates of the range of expected frequency falling between 1 and 3 kHz. The damping time, instead, cam be estimated as the ratio of the oscillation energy over the power emitted in gravity waves. Again, numerical results indicate a simple relation between the neutron star mass and radius (see e.g. Eq.~(8) of \citet{1998MNRAS.299.1059A}) which lead to damping times between 0.05 and 0.5 seconds (see also \citet{2005AIPC..751..211B}). We also note that an additional source of uncertainty stems from the conversion efficiency: for recent work on this aspect, see e.g. \citet{2011MNRAS.418..659L}. In summary, we estimate that for the quake driven glitches the uncertainty is about $72\%$, while for vortex driven glitches is of about $52\%$. 

The fractional uncertainties in pulsar distance range from $6\%$ up to $300\%$, with an average of $70\%$. This indicates that the dominant source of uncertainty among all experimental variables depends on the pulsar in question, as well as the model. For MQ and CQ, the radius will in most cases be the primary consideration, due to the large exponential dependence on $r$. In QO the frequency of oscillations will typically be the biggest source of uncertainty, while for FS and FQ the distance will be the dominant factor in all but the most tightly constrained cases.

\begin{figure}[t]
\centering
\includegraphics[scale=.6]{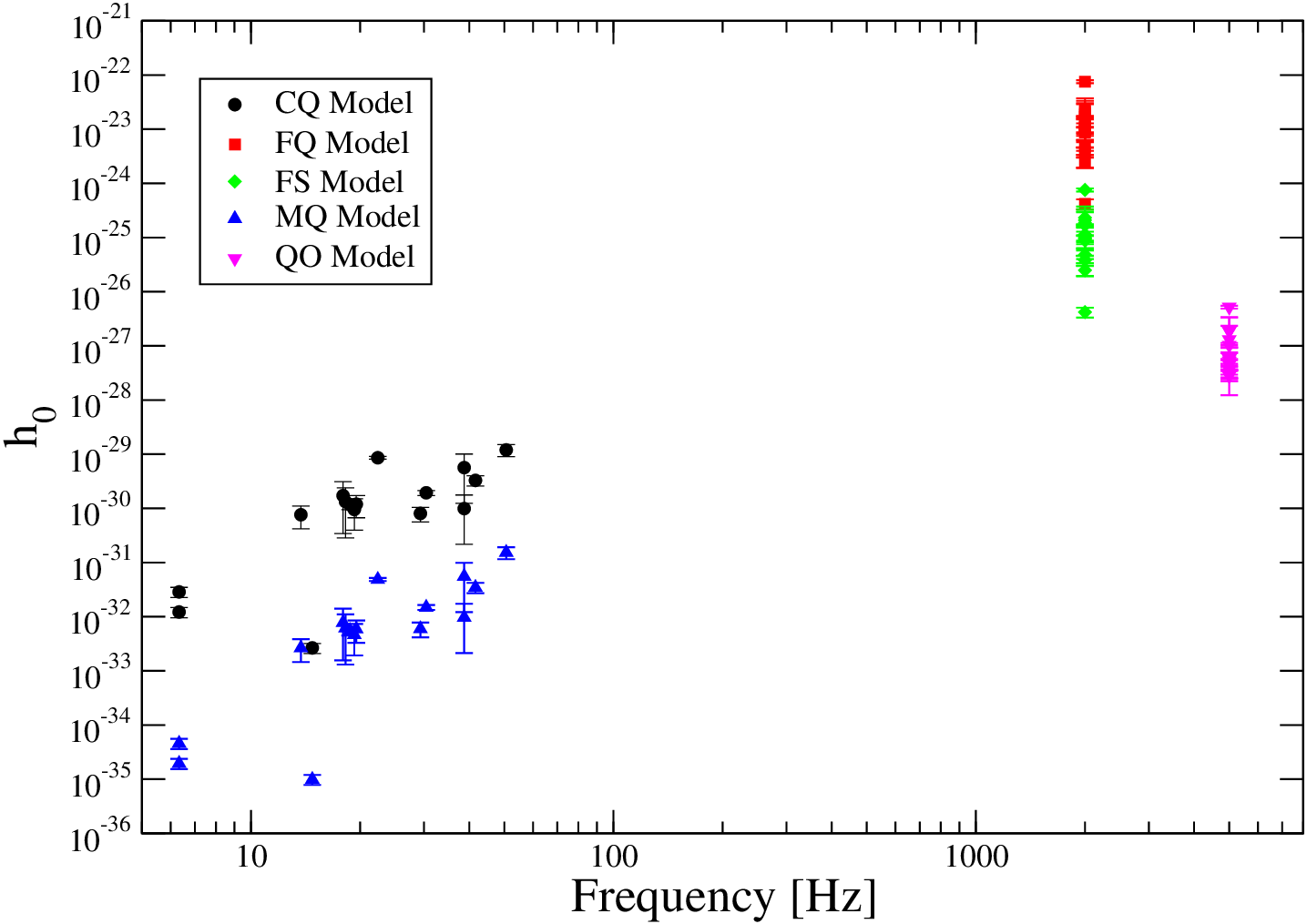}
\caption{Predictions for the peak gravity wave frequency and amplitude, for the five theoretical models we consider here.  Error-bars correspond to distance uncertainties only, see the text for a discussion of other sources of uncertainties.}\label{fig:freq}
\end{figure}

We show in \figref{freq}, in the form of data-points with error-bars, the predictions for the peak frequencies and dimensionless gravity wave amplitude for all 5 models and all pulsars, at central values of the distance. The error-bars shown in the figure only reflect the uncertainties associated with pulsar distances, as the systematic errors associated with each theoretical model are universal for each model and thus the same for all pulsars. 
 
For each of the pulsars and glitches in Table 2, we have the following general hierarchy among the amplitude of the predicted GW emission: $FQ > FS > QO > CQ > MQ$. It is not surprising that the models that predict the biggest signals are FQ and FS of Abadie et al., 2010, since in this case the energy of the glitch is assumed to be completely transferred to $f$-mode oscillations that are the strongest emitters of gravitational radiation.  
 
Note that pulsar J0835-4510, also known as ``Vela'', is one of the strongest predicted emitter across the latter three models (FQ, FS and QO) and the third strongest emitter for the first two models (MQ and CQ). This is expected, given that the magnitude of the glitch and pulsar distance will be the primary considerations for the signal in the latter three models where the other parameters are estimated and held constant among the pulsars. In contrast, J1952+3252 and J2229+6114 are predicted to have a larger gravity wave amplitude than Vela in the first two models (MQ and CQ) because such amplitude depends steeply on frequency (respectively, $h_0\sim\Omega^4$ and $\sim\Omega^2$), and these two pulsars have among the highest frequency in the set. 

\begin{figure}[t]
\centering
\includegraphics[scale=.6]{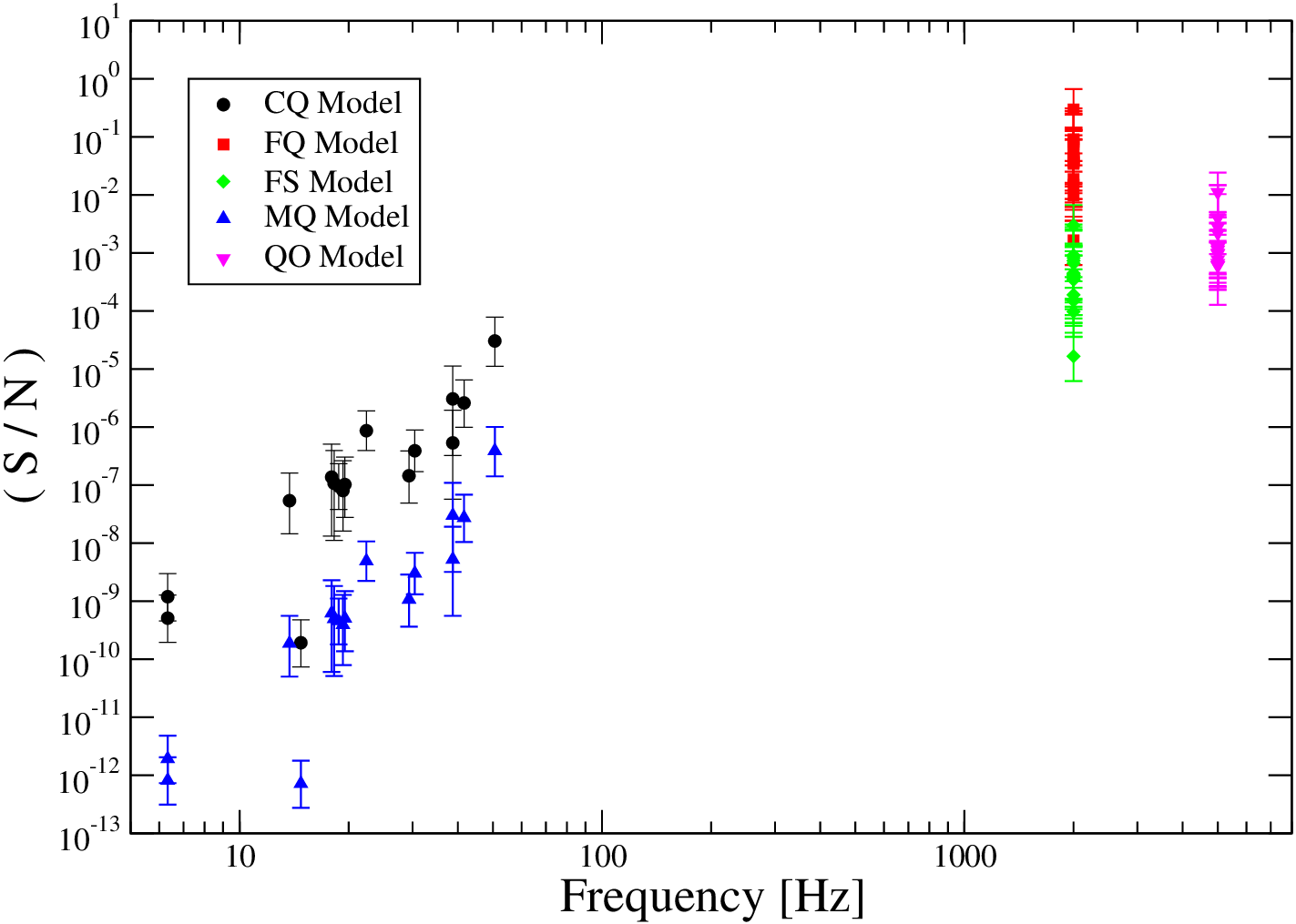}
\caption{Predictions for the signal to noise ratio (see text for details on how we estimate this quantity), for the five theoretical models we consider here.  Error-bars correspond to both the uncertainties on $h_0$ highlighted in fig.~\ref{fig:freq} and to the signal duration uncertainties.}\label{fig:detect}
\end{figure}
  
\figref{detect} presents an estimate of the signal to noise ratio for current-generation detectors (LIGO S6, \citet{2012arXiv1203.2674T}). We estimate the signal to noise ratio (S/N) as follows \citep{maggiore}:
\begin{equation}
({\rm S}/{\rm N})\simeq h_o \sqrt{\frac{T_{\rm obs}}{S_n}},
\end{equation}
where $S_n$ is the detector's noise spectral density, and where $$T_{\rm obs}={\rm min}\left({\rm observation\ time},\ {\rm source\ duration}\right).$$ As stated above, we assume the the observation time is always longer than the source duration for all models. We derive values for $S_n$ from \citet{2012arXiv1203.2674T}. We estimate the uncertainty associated with the gravity wave amplitude uncertainty and with the signal duration uncertainty by calculating the ``maximal'' (S/N) for the longest signal duration and the largest amplitude, and the ``minimal'' (S/N), vice versa, for the shortest signal duration and the smallest amplitude, for a given glitch and theoretical model. For reference, typical (S/N) values for detection for a coherent search are 11.4 for a coherent search, 30 for a semi-coherent search and of order 10 for burst searches \citep{maggiore}. Advanced LIGO will achieve an improvement of about one order of magnitude in the (S/N) sensitivity. 

We conclude that the largest signal-to-noise ratios are expected for the FQ and for the QO models. With central values for the pulsar distances, none of these models is expected to produce a large enough signal to be detectable with current or future detectors. The CQ and MQ models predict signal to noise values that are definitely well below any current or future detector sensitivity.
  
\begin{figure}[t]
\centering
\includegraphics[scale=.6]{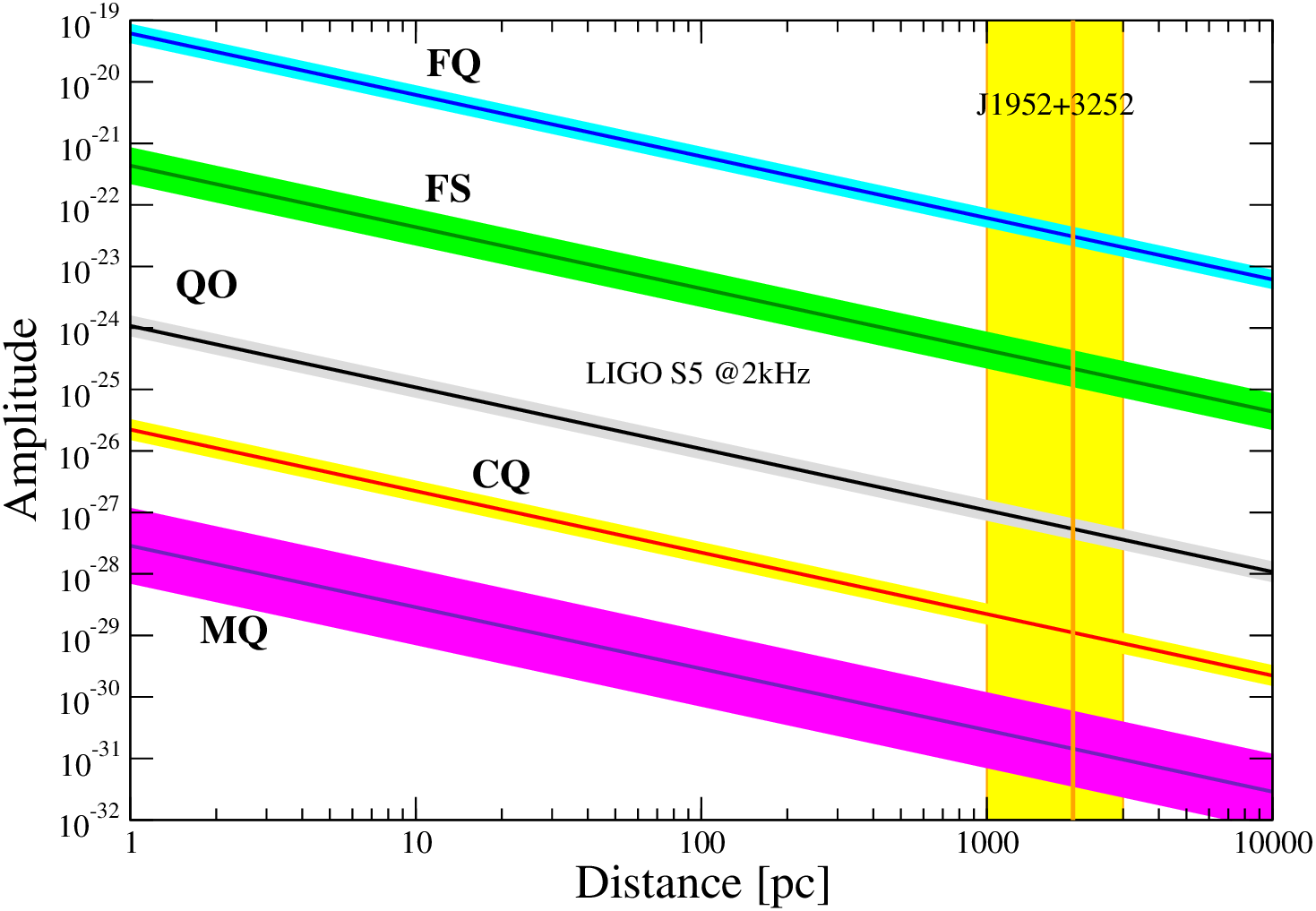}
\caption{The distance dependence of the amplitude for GW emission. We use for reference the glitch in J1952+3252, which aside from its distance is the most favorable glitch to GW emission in all models (with the exception of the QO model). The vertical yellow band indicates the distance estimate for J1952+3252 and the 2$\sigma$ uncertainty band.}\label{fig:dist}
\end{figure}

We show in \figref{dist} the predictions for the gravity wave amplitude for the five theoretical models we consider here, for the glitch in J1952+3252, as a function of the distance to the pulsar, in pc, which we treat here as a free parameter (we do indicate with a vertical band the expected distance range for this object). The bands indicate, and visualize, the theoretical uncertainties we estimated above. The hierarchy we display corresponds to the discussion outlined above. 
For reference, searches for burst-like signals from the Vela pulsar (which employed 120 seconds of data centered on the glitch epoch) resulted in limits in the range between $6.3\times 10^{-21}$ and $1.4\times 10^{-20}$ for the peak intrinsic strain amplitude. The vertical band indicates the 2$\sigma$ distance uncertainty for the pulsar distance, whose central value is 2 kpc (vertical orange line).


The expected (S/N) ratio for this pulsar would fall within the detectability range for current detectors for a pulsar distance  less than one order of magnitude closer than the central distance estimate shown in the plot for model FQ, and for a pulsar distance one-to-two orders of magnitude closer than the central distance for the FS and QO models. No detection is expected even for very small and unrealistic pulsar distances for the CQ or MQ models.


Our study indicates quite conclusively that gravitational wave signals predicted by the two models arising from Ekman flow, MQ and CQ, will likely be undetectable by the next generation of detectors. Models of gravitational waves emitted by quasi-radial oscillations might have very long signal durations and potentially could produce a detectable signal for some of the gamma-ray pulsar glitches under consideration. Predictions for the FS and FQ models generically fall outside the range of detectability of current detectors, but for certain pulsars, assuming the distance to the pulsar is significantly reduced compared to current estimates, detectability is not excluded. In all cases, signals are associated with a high peak frequency, likely on the order of a few kHz.

In particular, the most promising target glitches are associated, for both FS and FQ models, and in decreasing order, with pulsars J0835-4510 (Vela), J2229+6114, J1952+3252, J1413-6205, J1023-5746 and J0007+7303. The S/N predictions for all six gamma-ray pulsar glitches fall within a factor 2.5. All six pulsars have spin frequencies in the range between 3 and 25 Hz, and three of the pulsars are radio quiet while three are radio loud. Some of the most promising events stem from pulsars with very large distance uncertainties.

Of course, we may yet discover pulsar glitches corresponding to events that are closer and more energetic than those considered here. In addition, it is possible that the internal dynamics of some glitches may more effectively excite the necessary oscillations than those that we have observed thus far. 

It is important to note that the typical time localization for gamma-ray glitches is of about 10 minutes \cite{fermisymp}.  This is much worse than the accuracy with which glitches can be localized at radio frequencies (for example, the glitch considered in \citet{Abadie:2010sf} was localized to within 17 s). Estimating the computational cost associated with the search for a burst-like signal that is localized only to within 10 minutes lies beyond the scopes of the present study, but it will be an important consideration. Such estimate will also depend upon the search strategy \citep[see e.g.][]{Prix:2011qv}.

\section{Discussion and Conclusions}\label{sec:discussion}

 We used data on the first detected gamma-ray pulsar glitches, discovered with the Fermi Gamma-Ray Large Area Telescope, to predict gravity wave signals in the context of five different theoretical models. Gamma-ray glitches are promising events to be associated with a large gravity wave signal: for a glitch to be visible in gamma rays (as opposed to radio) the pulsar must be both nearby enough and the glitch large enough to trigger a detection with the sparse set of detectable gamma-ray events.
 
In this study, we employed five well-motivated (although in some cases simplistic) theoretical models to estimate the gravity wave signal given the input glitch parameters, including the glitch frequency change, the pulsar frequency and frequency derivative, the change in frequency derivative and the pulsar distance. We attempted as detailed an estimate of the systematic uncertainties for each model as possible, and concluded that often the pulsar distance is one of key unknowns driving the largest uncertainty associated with the anticipated gravity wave signal (although for certain models the distance error is subdominant to other sources of systematic uncertainty).

We estimated detectability based on signal-to-noise ratio, and found that for two models, specifically for the quake-driven $f$-mode oscillations model (FQ) and for the superfluid driven $f$-mode oscillations model (FS), the estimated gravity wave signal lies below, but not drastically distant, from the performance of current detectors, for central values of the pulsar distances. The same is true for the quasi-oscillation QO model. Should some of the glitching gamma-ray pulsars be closer than current estimates, the FS, FQ and QO models predict promising signals that would warrant a dedicated search.

We stress that many of the glitch events we considered here correspond to radio-quiet objects (see Table I), and have thus not been searched for in past LIGO data. We believe it would be exciting to look for a signal from those events in archival LIGO data. Looking at the future, this study emphasizes the importance of the connection between the violent universe as seen by gamma-ray observatories and the quest for gravity waves. This connection might well lead to exciting discoveries with current or future gravity wave observatories.

\acknowledgments
We thank Mario Belfiore and Michael Dormody for many illuminating discussions and clarifications, and Steve Ritz for discussions and input in the early stages of this work. We are also very thankful to the Referee for several insightful suggestions, including the discussion of the signal to noise, that greatly improved this manuscript. SP is partly supported by the US Department of Energy under Contract DE-FG02-04ER41268.

\bibliography{draft20}
\clearpage

\begin{sidewaystable}
\scriptsize
           \begin{tabular}{| l | c | c | c | c | c | c | c |} \toprule[.03em] 
		\toprule 
	{\textsc{Pulsar}}  & Distance  (kpc) & $f$ (hz) &  $\dot f$ (hz/s) & $\frac{\Delta f}{f}$ & Glitch Time (MJD) &  Radio? \\
	\hline \hline
	\midrule
         
         J0007+7303   &                      $1.4 \pm .3$ &        $ 3.1658268380$&       $-3.58\times10^{-12}$&        $5.54\times10^{-7}$&    $54954$       &  Quiet \\ \hline				
	
	J0007+7303   &		    $1.4 \pm .3$ &       $ 3.1658268380$ &      $-3.58\times10^{-12}$ &        $1.26\times10^{-6}$& $55466$ & Quiet \\ \hline
	
	J0205+6449   &		        $2.6-3.2$  &      $15.2143754050$ &      $-4.48\times10^{-11}$ &        $1.74\times10^{-6}$& $54795$ & Loud \\	\hline									
								
	J0835-4510    &    $.287^{+.019}_{-.017}$&        $11.189512263$ &       $-1.56\times10^{-11}$&      $1.92\times10^{-6}$& $55408$ & Loud \\ \hline
						
	J1023-5746    &                                 $2.4*$ &        $8.9703354320$ &       $-3.08\times10^{-11}$&       $3.56\times10^{-6}$& $55041$ & Quiet \\ \hline
						
         J1124-5916     &           $4.8^{+.7}_{-1.2} $ &          $7.379534279$ &          $-4.1\times10^{-11}$&      $3.06\times10^{-8}$& $55191$ & Faint \\ \hline
	          			
	J1413-6205     &                               $1.4*$  &        $9.1123866570$ &        $-2.29\times10^{-12}$&      $1.73\times10^{-6}$& $54735$ & Quiet \\ \hline
	          			
	J1420-6048     &                   $5.6 \pm 1.7$ &            $14.66123111$& $-1.78173\times10^{-11}$&       $1.35\times10^{-6}$& $55435$ & Loud \\ \hline
	
	J1709-4429     &                          $1.4-3.6 $ &        $9.7564601870$ &     $  -9.03\times10^{-12}$ &        $2.75\times10^{-6}$&   $54693$ & Loud  \\ \hline
	              		
	J1813-1246     &                  $3.19 \pm .69$ &          $20.80188547$ &        $-7.61\times10^{-12}$&                       $1.16\times10^{-6}$& $55094$ & Quiet \\ \hline
	
	J1838-0537     &            $1.49\text{---}3.93$&                        $6.863$&        $2.1896\times10^{-11}$&         $5.5\times10^{-6}$& $55100$ &  Quiet\\ \hline
						
	J1907+0602    &                       $3.2 \pm .6$ &          $9.377661105$ &      $-7.631\times10^{-12}$&        $4.66\times10^{-6}$& $55422$ & Faint \\ \hline
						
	J1952+3252    &                      $2.0 \pm .5 $ &          $25.29471117$ &      $-3.727\times10^{-12}$&          $1.5\times10^{-6}$ & $55325$ & Loud \\ \hline
					
         J2021+3651     &          $2.1^{+2.1}_{-1.0}$ &        $9.6390889420$ &       $-8.87\times10^{-12}$ &         $2.23\times10^{-6}$& $55109$ & Loud \\ \hline
         
         J2229+6114     &                              $.8-6.5 $&         $19.361874486$&  $-2.91834\times10^{-11}$&        $2.05\times10^{-7}$&  $55130$ &   Loud    \\ \hline                      
        
         J2229+6114     &                             $.8-6.5 $ &        $19.361874486$ &  $-2.91834\times10^{-11}$&       $1.231\times10^{-6}$& $55599$ & Loud \\ \hline

\bottomrule
\end{tabular} 
\begin{center}
\begin{minipage}{7in}
	\caption{Distances, in kpc, to the 14 gamma-ray pulsars we consider here, as well as their rotation and glitch parameters. The distances are consolidated from \citet{1475-7516-2010-02-016}, \citet{0067-0049-187-2-460}, \citet{0004-637X-725-1-571}, and \citet{1674-4527-11-7-007}, while for all other entries we utilize data presented in \citet{fermisymp}, \citet{2011AAS...21743405D}, and Pletsch et al. \cite{2041-8205-755-1-L20}. $f$ indicates the frequency of rotation of the pulsar, $\dot f$ is the spin-down rate, $\frac{\Delta f}{f}$ is the fractional change in frequency during the glitch. Note that there are two gamma-ray glitches detected for pulsars J0007+7303 and J2229+6114. Distance values with an asterisk may be larger or smaller than those given by a factor of 2 or 3. The last column indicates if the pulsar is radio loud, quiet or faint.}
	\end{minipage}
	\end{center} 
\end{sidewaystable}

\begin{sidewaystable}[t!]
\scriptsize
\scalebox{1}{
           \begin{tabular}{| l | c | c | c | c | c | c | c |} \toprule[.03em] 
		\toprule 
	{\textsc{Pulsar}}  & MQ  & CQ & QO & FS & FQ & $f_{+}\text{ and }f_{\times\text{-}pole}$\;\; (hz) & $f_{\times\text{-}equator}$ \;\; (hz)\\
	\hline \hline
	\midrule
	J0007+7303    &		                             $2\times10^{-35}$&                                                            $10^{-32}$ &                      $2\times^{-28}\text{---}3\times10^{-28}$ &    $2\times10^{-26}\text{---}3\times10^{-26}$&  $2\times10^{-24}\text{---}3\times10^

	{-24}$ & 6.33 & 3.17 \\ \hline

	J0007+7303     &   $4\times10^{-35}\text{---}6\times10^{-35}$&       $2\times10^{-32}\text{---}3\times10^{-32}$&     $2\times10^{-28}\text{---}3\times10^{-28}$&    $3\times10^{-26}\text{---}5\times10^{-26}$&  $3\times10^{-24}\text{---}5\times10^
	
	{-24}$ & 6.33 & 3.17 \\ \hline
										
	J0205+6449   &		 $10^{-32}\text{---}2\times10^{-32}$&                                                $2\times10^{-30}$&                    $9\times10^{-28}\text{---}10^{-27}$&                  $9\times10^{-26}\text{---}10^{-25}$&  $9\times10^{-24}\text{---}10^{-23}$& 30.43& 15.21 \\	\hline									
								
	J0835-4510    &                                                $5\times10^{-32}$&        $8\times10^{-30}\text{---}9\times10^{-30}$&      $5\times10^{-27}\text{---}6\times10^{-27}$&      $7\times10^{-25}\text{---}8\times10^{-25}$&  $7\times10^{-23}\text{---}8\times10^{-23}$ & 22.38 & 11.19 \\ \hline
						
	J1023-5746   &                    $2\times10^{-33}\text{---}10^{-32}$&        $3\times10^{-31}\text{---}3\times10^{-30}$&     $3\times10^{-28}\text{---}2\times10^{-27}$&      $3\times10^{-26}\text{---}3\times10^{-25}$&                              $3\times10^{-24}\text{---}3\times10^{-23}$ & 17.94 & 8.97 \\ \hline
						
	J1124-5916   &                     $8\times10^{-36}\text{---}10^{-35}$&       $2\times10^{-33}\text{---}3\times10^{-33}$&      $4\times10^{-28}\text{---}5\times10^{-28}$&       $3\times10^{-27}\text{---}5\times10^{-27}$&                 $3\times10^{-25}\text{---}5\times10^{-25}$ & 14.76 & 7.38 \\ \hline
	 
	 J1413-6205  &                                  $10^{-33}\text{---}10^{-32}$&       $3\times10^{-31}\text{---}2\times10^{-30}$&                                $10^{-28}\text{---}10^{-27}$&       $4\times10^{-26}\text{---}3\times10^{-25}$&                 $4\times10^{-24}\text{---}3\times10^{-23}$ & 18.22 & 9.11 \\ \hline
	          			
	J1420-6048   &         $4\times10^{-33}\text{---}8\times10^{-33}$&                   $6\times10^{-31}\text{---}10^{-30}$&       $2\times10^{-28}\text{---}5\times10^{-28}$&       $3\times10^{-26}\text{---}6\times10^{-26}$&                 $3\times10^{-24}\text{---}6\times10^{-24}$ & 29.32 & 14.66 \\ \hline
	           			
	 J1709-4429  &         $3\times10^{-33}\text{---}8\times10^{-33}$&      $7\times10^{-31}\text{---}2\times10^{-30}$&      $3\times10^{-28}\text{---}8\times10^{-28}$&       $6\times10^{-26}\text{---}2\times10^{-25}$&                 $6\times10^{-24}\text{---}2\times10^{-23}$ & 
	 
	 19.51 & 9.76 \\ \hline
	              		
	J1813-1246    &         $3\times10^{-32}\text{---}4\times10^{-32}$&      $3\times10^{-30}\text{---}4\times10^{-30}$&      $4\times10^{-28}\text{---}6\times10^{-28}$&                     $8\times10^{-26}\text{---}10^{-25}$&                               $8\times10^{-24}\text{---}10^{-23}$& 41.6 & 20.80 \\ \hline

	J1838-0537    &                             $10^{-33}\text{---}4\times10^{-33}$&               $4\times10^{-31}\text{---}10^{-30}$&     $4\times10^{-28}\text{---}9\times10^{-28}$&         $6\times10^{-26}\text{---}2\times10^{-25}$&                           $6\times10^{-24}\text{---}2\times10^{-23}$                                                                   & 13.76         &   6.86  \\ \hline

	J1907+0602   &         $4\times10^{-33}\text{---}7\times10^{-33}$&         $9\times10^{-31}\text{---}\times10^{-30}$&      $3\times10^{-28}\text{---}4\times10^{-28}$&                       $8\times10^{-26}\text{---}10^{-25}$&               $8\times10^{-24}\text{---}10^{-23}$ & 18.76 & 9.38 \\ \hline
						
	J1952+3252   &                      $10^{-31}\text{---}2\times10^{-31}$&                    $9\times10^{-30}\text{---}10^{-29}$&       $4\times10^{-28}\text{---}7\times10^{-28}$&         $2\times10^{-25}\text{---}3\times10^{-25}$&                $2\times10^{-23}\text{---}3\times10^{-23}$ & 50.59 & 25.29 \\ \hline
					 
         J2021+3651    &         $2\times10^{-33}\text{---}7\times10^{-33}$&      $4\times10^{-31}\text{---}2\times10^{-30}$&        $2\times10^{-28}\text{---}9\times10^{-28}$&          $5\times10^{-26}\text{---}2\times10^{-25}$&                             $5\times10^{-24}\text{---}2\times10^{-23}$ & 19.28 & 9.64 \\ \hline
                                                       
         J2229+6114     &        $2\times10^{-33}\text{---}2\times10^{-32}$&      $2\times10^{-31}\text{---}2\times10^{-30}$&        $4\times10^{-28}\text{---}3\times10^{-27}$&          $2\times10^{-26}\text{---}2\times10^{-25}$& $2\times10^{-24}\text{---}2\times10^{-23}$& 
         
         38.73 & 19.36 \\ \hline

         J2229+6114      &                  $    10^{-32}\text{---}\times10^{-31}$ &                                $10^{-30}\text{---}10^{-29}$&          $4\times10^{-28}\text{---}3\times10^{-27}$&          $5\times10^{-26}\text{---}4\times10^{-25}$& $5\times10^{-24}\text{---}4\times10^{-23}$  & 38.73 & 19.36 \\ \hline
\bottomrule
\end{tabular}
} 
\begin{center}
\begin{minipage}{7in}
	\caption{Predicted amplitudes for each of the pulsars and models.  The ranges are due to uncertainty in the distances to the pulsars. CQ refers to the Current Quadrupole prediction, $MQ=$ Mass Quadrupole, $QO=$ Quasi-radial Oscillations, $FQ=$ Quake Driven $f$-Mode Oscillations, and $FS=$ Superfluid Driven $f$-Mode Oscillations.The last two columns contain the frequencies of the gravitational wave signals as predicted by the CQ and MQ models, for each of the polarizations observed along the line of sight of the pole and equator. In general there will be signals at both frequencies, with the relative contribution of each signal depending on the inclination of the pulsar.}
	\end{minipage}
	\end{center} 
\end{sidewaystable}

\end{document}